\documentclass[prb,twocolumn,showpacs,preprintnumbers,superscriptaddress,amsmath,amssymb]{revtex4-1}
\usepackage{graphicx}
\usepackage{dcolumn}
\usepackage{bm}
\usepackage{amsfonts}
\usepackage{amsmath}
\usepackage{amssymb}
\usepackage{xcolor}
\usepackage{enumerate}
\usepackage[utf8x]{inputenc}

\newcommand{\rucl}{$\alpha$-RuCl$_3$}
\newcommand{\oscl}{OsCl$_3$}

\begin{document}

\title{Magnetic anisotropy energies and metal-insulator transitions in monolayers of $\alpha$-RuCl$_3$ and OsCl$_3$ on graphene}

\author{P. H. Souza}
\affiliation{Instituto de F\'isica, Universidade Federal de Uberl\^andia,   C.P. 593, 38400-902, Uberl\^andia, MG, Brazil}
\author{D. P. de Andrade Deus}
\affiliation{Instituto Federal de Educa\c{c}\~ao, Ci\^{e}ncia e Tecnologia de Goi\'as, 
Departamento de \'Areas Acad\^{e}micas, Jata\'i, GO, Brazil}
\author{W. H. Brito}
\email{walber@fisica.ufmg.br}
\affiliation{Departamento de  F\'{\i}sica, Universidade  Federal de Minas Gerais, C. P. 702, 30123-970, Belo Horizonte, MG, Brazil}
\author{R. H. Miwa}
\affiliation{Instituto de F\'isica, Universidade Federal de Uberl\^andia,   C.P. 593, 38400-902, Uberl\^andia, MG, Brazil}

\date{\today}
    
\begin{abstract}
Transition metal thriclorides, with $4d$ or $5d$ electrons, are materials at the forefront of recent studies about the interplay of spin-orbit coupling and strong Coulomb interactions. Within our first-principles calculations (DFT+$U$+SOC) we study the effects of graphene on the electronic and magnetic properties of the monolayers of $\alpha$-RuCl$_3$ and OsCl$_3$. Despite the spatially inhomogeneous $n$-type doping induced by graphene, we show that the occupancy of the upper Hubbard bands of MLs of \rucl and OsCl$_3$ can be tuned through external electric fields, and allows the control of (i) metal-insulator transitions, and (ii) the magnetic easy-axis and anisotropy energies. Our findings point towards the tunning of electronic and magnetic properties of transition metal thriclorides monolayers by using graphene and external electronic fields.

\end{abstract}

\maketitle

\section{Introduction}

Magnetic interactions in correlated materials are well known for giving rise to new energy scales and emerging properties. In  Mott insulators, with localized $d$-electrons, these interactions can give rise to an antiferromagnetic ground state,~\cite{mott,andersonSe} which upon hole doping leads to unconventional superconductivity.~\cite{patrickLee_SC} On the other hand, in $f$-electron materials the antiferromagnetic interaction between local moments and the conduction electrons leads to the Kondo effect and the appearance of strongly renormalized quasiparticles.~\cite{coleman_hf} More recently, the interplay of spin-orbit coupling (SOC), strong Coulomb and the emerging magnetic interactions has become of great interest since it can give rise to unusual electronic phases and nontrivial topology.~\cite{soc_str_r1,soc_str_r2}

At the forefront of recent studies in this field are the 4$d$ and 5$d$ based materials, such as ruthenium and osmium compounds. In special, $\alpha$-RuCl$_3$ has attracted great interested since it is a candidate for the realization of a Kitaev-like quantum spin liquid.~\cite{sandilands_PRL} However, this material is found to be a spin-orbit Mott insulator which orders antiferromagnetically at low temperatures.~\cite{plumb_PRB} The intriguing magnetic interactions in $\alpha$-RuCl$_3$ are also responsible for competing magnetic phases. More recently, several works have addressed the strength and nature of these interactions. For instance, recent X-ray scattering data have obtained a ferromagnetic Kitaev term in $\alpha$-RuCl$_3$,~\cite{xray_NatPhys} in good agreement with previous calculations.~\cite{winter_NatComn, maksimov_PRR}
Moreover, within density functional theory (DFT)+$U$+SOC calculations, Kim {\it et al.}~\cite{kim2PRB015} found that a zigzag antiferromagnetic (ZZ-AFM) phase of $\alpha$ -RuCl$_3$ is slightly more stable than the FM configuration for $U$ values between 1.0 and 3.5\,eV, where the former is characterized by an energy gap of $\sim$0.8 eV. Such a nearly degeneracy between FM and ZZ-AFM phases has been confirmed by recent works.~\cite{huangPRB2017,tianNanoLett2019}.

The monolayers (MLs) of the 4$d$ and 5$d$ transition metal thriclorides have also attracted great attention due to their magnetic and electronic properties. In the work of Huang {\it et al.},~\cite{huangPRB2017} the authors explored the quantum anomalous Hall (QAH) effect in monolayer of $\alpha$-RuCl$_3$. However, instead of a semiconducting phase,~\cite{kim2PRB015} they found a semimetallic system even upon the inclusion of the SOC. In contrast, the semiconducting character of the ML $\alpha$-RuCl$_3$ has been confirmed by recent calculations of Sarikurt {\it et al.}\,~\cite{sarikurtPCCP2018}, with energy gaps of about 0.7~eV (FM) and 1.0\,eV  (ZZ-AFM) for $U$=2.0\,eV. From the experimental side, neutron scattering experiments ~\cite{searsPRB2015} indicated that single crystals of $\alpha$-RuCl$_3$, with Ru$^{3+}$ (4$d^5$) ions exhibit a ZZ-AFM configuration at low temperatures.

Similar magnetic phases have also been explored in the ML OsCl$_3$, which in contrast to $\alpha$-RuCl$_3$, is predicted to be ferromagnetic.~\cite{sheng_prb} In addition, ML OsCl$_3$ has been considered as a candidate of a quantum anomalous Hall (QAH) insulator,~\cite{sheng_prb} although the Coulomb interactions counteract and favor the appearance of a Mott insulating phase. More recently, it was pointed out that the parent compound Os$_{0.55}$Cl$_2$ presents features of a quantum spin-liquid, with gapless magnetic fluctuations which prevent any magnetic ordering at low temperatures.~\cite{McGuirePRB}
Besides the feasibility of layered systems,~\cite{gronkePCCP2018} and the prediction of Hubbard-$U$ dependent QAH phase in the ML of OsCl$_3$,~\cite{sheng_prb} there are few works exploring its electronic and magnetic properties. 

Another interesting property of these two-dimensional (2D) materials is their magnetic anisotropy energy (MAE). In the Ref.~\onlinecite{sarikurtPCCP2018}, the authors reported a MAE of $\sim$18.8\,meV with the easy axis parallel to the $\alpha$-RuCl$_3$\ surface. Further calculations confirmed the energetic preference for the in-plane magnetization, however they obtained a MAE of 0.80\,meV/Ru-atom for the ZZ-AFM phase,~\cite{iyikanatJMatChemC2018} which is much smaller than that obtained in Ref.~\onlinecite{sarikurtPCCP2018}. 

Previous reports have also explored different ways to modify the properties of these 2D magnets. In the case of $\alpha$-RuCl$_3$, strain and optically driven charge excitations can be used to induce magnetic phase transitions.~\cite{iyikanatJMatChemC2018,tianNanoLett2019,kaib_arxiv} Another interesting way is by interfacial interaction with graphene. In particular, it has been found that the interaction between graphene and ML $\alpha$-RuCl$_3$~\cite{Valenti_PRL} can enhance considerably the Kitaev interactions in latter, whereas it also modifies the Fermi surface topology of the former.~\cite{tianNanoLett2019} Rizzo {\it et al.}~\cite{rizzo_NanoLett}, for instance, found that interlayer charge transfer between the ML $\alpha$-RuCl$_3$ and graphene gives rise to plasmon polaritons with high mobility.

In this work, we study the the electronic and magnetic properties of the MLs of \rucl\, and \oscl\ in presence of graphene (Gr); \textit{i.e.} \rucl/Gr and \oscl/Gr heterostructures. We find a net charge transfer (of the order of 10$^{13}$/cm$^2$) from graphene to the transition metal trichlorides, however inhomogeneously distributed on the \rucl\, and \oscl\, surfaces.  The upper Hubbard bands of both $\alpha$-RuCl$_3$ and OsCl$_3$ become partially occupied, giving rise to correlated metallic phases.  More important, we demonstrate that graphene and external electric fields (EEFs) can be used to control the Ru-4$d$ (Os-5$d$) occupancy, and as a result,  induce (i) metal-insulator transitions,  and (ii) tune MAE in both two dimensional magnets, where the graphene monolayer acts as reservoir for the electron/hole doping of the MLs. For instance, in the case of \rucl/Gr, we observe a change of easy axis (from out of plane to in plane) accompanied by an enhancement of the MAE by a factor of six under external electric fields of 0.2 eV/\AA{}. These findings suggest that the occupancies of MLs $\alpha$-RuCl$_3$ and OsCl$_3$ Hubbard bands are essential to control the magnetization easy-axis and the corresponding magnetic anisotropic energies. This control can be achieved with the presence of graphene and external electric fields.

\section{\label{sec:method} Computational details}
Our density functional theory calculations were performed within the Perdew-Burke-Ernzehof generalized gradient approximation (PBE-GGA),~\cite{PBE} using projector augmented wave (PAW) potentials,~\cite{paw} as implemented in the Vienna ab initio Simulation Package (VASP) code.~\cite{vasp1,vasp2} The structural optimizations were done including van der Walls corrections (vdW-DF)~\cite{vdw_vasp} until the forces on each atom were less than 0.01 eV/\AA. A plane wave cutoff of 500 eV was used, and $k$-point meshes of ($15\times 15 \times1$), and ($31 \times 31 \times1$) for total energy and projected density of states calculations, respectively. The \rucl\, and \oscl\, monolayers were simulated using a slab with $\sqrt 3\times\sqrt 3$ surface periodicity, which corresponds to eight Ru/Os-atoms per unit cell [Fig.\,\ref{fig:FigMagnetic}(a)], and a vacuum region of 20\,\AA, separating a slab from its periodic images.

To treat the strong correlations related to electrons within Ru-4$d$ (Os-5$d$) states, we employed the DFT+$U$ functional of Dudarev,~\cite{dudarev} which takes into account the electronic interactions at a mean-field level by means of an effective partially screened Coulomb interaction $U_\text{eff}$. Likewise Refs.~\onlinecite{Valenti_PRL} and ~\onlinecite{sheng_prb}, we used values of $U_\text{eff} = U-J =$ 1.5 eV and 1.0 eV for MLs of $\alpha$-RuCl$_3$ and OsCl$_3$, respectively. The spin-orbit coupling (SOC) was also taken into account in our calculations.

\section{Results and Discussions}

\subsection{Electronic and magnetic properties of MLs $\alpha$-RuCl$_3$ and OsCl$_3$ within DFT+$U$+SOC}

\begin{figure}
\centering
 \includegraphics[width=\columnwidth]{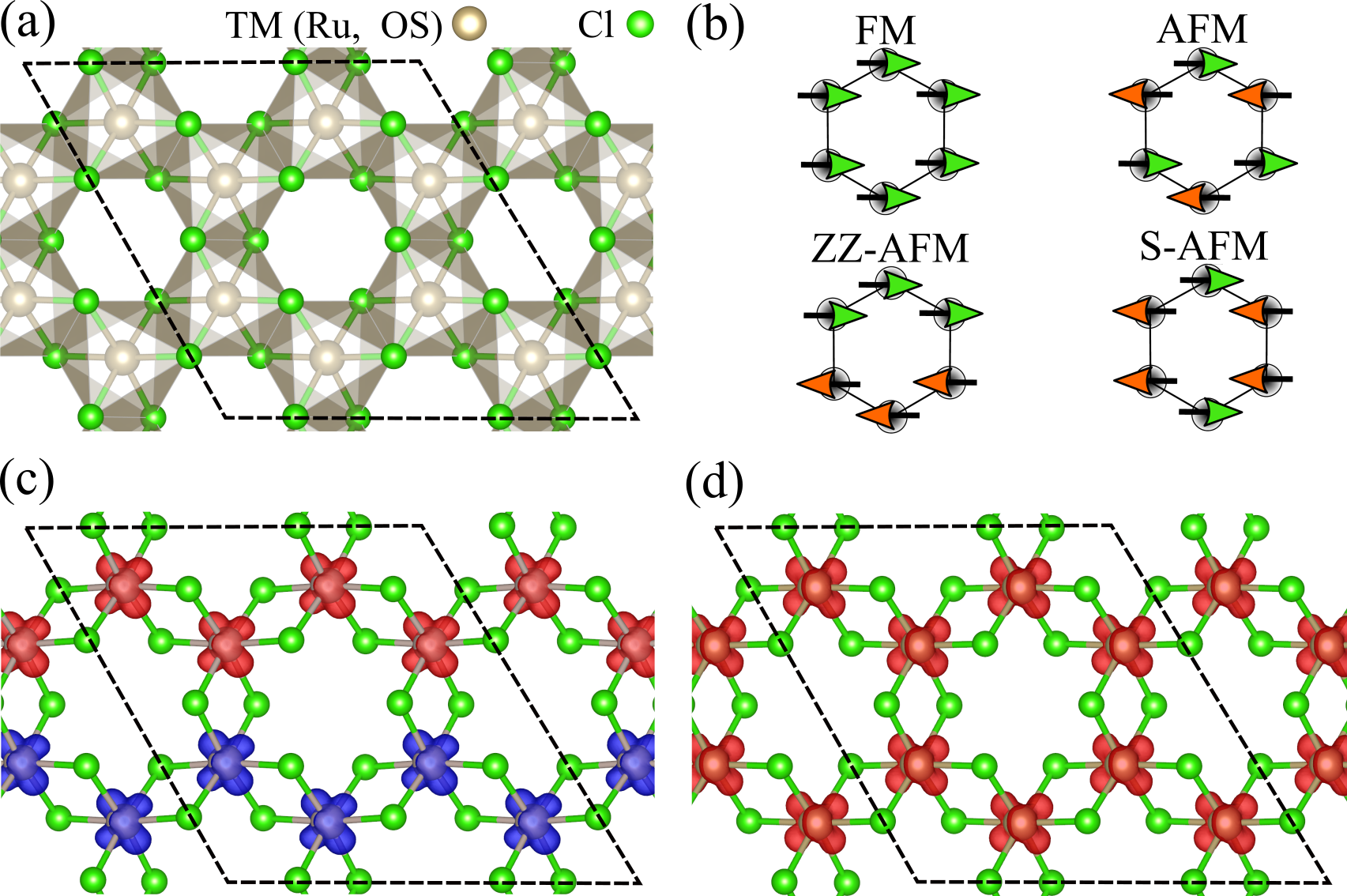}
 \caption{(a) Layered honeycomb structures of the monolayers of $\alpha$-RuCl$_3$ and OsCl$_3$. (b) Distinct magnetic configurations associated with Ru(Os) local moments considered in our work. In (c) and (d) we show the spin density distributions corresponding to ZZ-AFM and FM phase of $\alpha$-RuCl$_3$ and OsCl$_3$, respectively. The red (blue) isosurfaces correspond to spin up (down) charge densities.}
 \label{fig:FigMagnetic}
\end{figure}

The MLs of $\alpha$-RuCl$_3$ and OsCl$_3$ have a two-dimensional honeycomb like structure, with slightly distorted $TM$Cl$_6$ ($TM$\,=\,Ru, and Os) edge-sharing octahedra, as shown in Fig.~\ref{fig:FigMagnetic}(a). The corresponding crystal field splitting and occupancy of the Ru-4$d$ (Os-5$d$) states lead to well defined local moments, which can order into distinct magnetic configurations. Within our DFT+$U$+SOC approximation we investigate the relative energetic stability of several magnetic phases, \textit{viz.}: ferromagnetic (FM), Néel antiferromagnetic (AFM), zigzag antiferromagnetic (ZZ-AFM), and stripy-antiferromagnetic (S-AFM), which are schematically illustrated in Fig.~\ref{fig:FigMagnetic}(b).

In Table~\ref{table:magnec_pr} we present the obtained total energy differences ($\Delta E$) with respect to the lowest energy configuration, net magnetic moments ($M$), and band gaps ($E_{\text gap}$) of each magnetic phase. We find that ML $\alpha$-RuCl$_3$ has a semiconducting ZZ-AFM ground state. The corresponding spin density, $\Delta \rho^\text{spin} = \rho_\text{up}-\rho_\text{down}$, shows a $d_{xy}$ shaped $\Delta \rho^\text{spin}$ for the ZZ-AFM phase [Fig.~\ref{fig:FigMagnetic}(c)]. The local moment of $\approx$ 0.9 $\mu_{B}$/Ru atom is in good agreement with previous reports,~\cite{sarikurtPCCP2018,iyikanatJMatChemC2018} though, we obtain a distinct ground state magnetic configuration. From our calculations we find that ZZ-AFM is 4.5 meV/Ru-atom more stable than the FM state. In contrast, in the work of Sarikurt {\it et al.},~\cite{sarikurtPCCP2018} the FM state was found to be the ground state, with 20 meV/Ru-atom more stable than ZZ-AFM configuration. Further, Iyikanat {\it et al.} pointed out the ZZ-AFM configuration as the ground state configuration. According to these authors the energy difference between ZZ-AFM and FM configuration is around 0.8 meV/Ru-atom. This nearly degeneracy of ZZ-AFM and FM configurations was also reported in Refs.\,\onlinecite{kim2PRB015,tianNanoLett2019}.

Moreover, the stripy (S-AFM) and Néel (AFM) configurations are around 9 and 12 meV/Ru less stable than the ZZ-AFM configuration, respectively. These small energy differences obtained within our approach reflect the competing energy scales presented in ML $\alpha$-RuCl$_3$. In fact, as pointed out by Winter \textit{et al.}~\cite{winter_prb2016} the rich phase diagram of $\alpha$-RuCl$_3$ is ruled by competing Coulomb, kinetic and spin-orbit energy scales as well as long-range interactions, where the latter are very sensitive to the structural details. It turns out that ZZ-AFM ground state can be attributed to the presence of a large third-neighbor Heisenberg coupling between the Ru$^{3+}$ local moments. Although the monolayer of $\alpha$-RuCl$_3$ exhibits nearly degenerated magnetic configurations, the insulating nature of the system is preserved. In particular, the antiferromagnetic configurations have larger band gaps. For instance, the ZZ-AFM $\alpha$-RuCl$_3$ has a band gap of around 200 meV larger than the FM configuration. 


We next evaluate the magnetic anisotropy energy (MAE), which is  defined as the total energy ($E$) difference of the \rucl, and \oscl\, systems with the magnetization parallel ($\parallel$) and perpendicular ($\perp$) to the slab surface, 
$$
\text{MAE} = E_\parallel - E_\perp.
$$
Here, the total energies $E_\parallel$ and $E_\perp$ were obtained by using the force theorem.~\cite{weinerPRB1985total,wangJMMat1996validity,yangPRB2017strain}  We found that the ZZ-AFM $\alpha$-RuCl$_3$ presents an energetic preference for the out-of-plane ($\perp$) magnetization with MAE\,=\,0.163\,meV/Ru-atom. Meanwhile previous studies pointed out an energetic preference for the in-plane ($\parallel$) magnetization by 18.88\,meV,\cite{sarikurtPCCP2018} and 0.80\,meV/Ru-atom\cite{iyikanatJMatChemC2018} for the FM and ZZ-AFM configurations, respectively.

As discussed above, the equilibrium geometry plays an important role in the magnetic properties of $\alpha$-RuCl$_3$. Indeed, we found a subtle commitment between the atomic positions and the  preferential orientation of the magnetic moment. We examined six slightly different equilibrium geometries obtained upon small perturbations on the initial (starting) atomic positions, and spin-configurations. We found that (i) the $\alpha$-RuCl$_3$ systems with energetic preference for in-plane magnetization (MAE\,$<$\,0) present Ru-Cl bonds lengths ($d_\text{Ru-Cl}$) with a nearly uniform distribution, characterized by a deviation ($\sigma_{d_\text{Ru-Cl}}$) of about 0.006\,\AA, whereas (ii) the ones with MAE\,$>$\,0 present less uniform $d_\text{Ru-Cl}$ distribution, with $\sigma_{d_\text{Ru-Cl}}\approx\,0.017$\,\AA. In all cases, we found lowest energies in (ii) by $\sim$\,30\,meV/Ru-atom when compared with (i). Thus,  providing further support to the energetic preference for the out-of-plane magnetization in \rucl.

\begin{table}[!htb]
\caption{Magnetic moments $M$ ($\mu_\text{B}$/$TM$-atom), relative energetic stabilities $\Delta E$ (meV/$TM$-atom), and band gaps E$_\text{gap}$ (eV) of distinct magnetic configurations of $\alpha$-RuCl$_3$ and OsCl$_3$ monolayers.}
\begin{ruledtabular}
\begin{tabular}{cccc}
  & $M$  & $\Delta E$ & E$_\text{gap}$ \\ \hline
             &     \multicolumn{2}{c}{$\alpha$-RuCl$_3$}   &        \\ 
ZZ-AFM &   0.89 &  0.0 &  0.74   \\
FM &  0.90        &  4.5 &  0.50   \\  
S-AFM &  0.90        &  8.6 &  0.66   \\
AFM & 0.86          & 11.6 & 0.80 \\ \hline 
             &     \multicolumn{2}{c}{OsCl$_3$}   &        \\
 ZZ-AFM &  0.80        &  3.0 &  0.77   \\
FM &  0.92        &  0.0 &  0.54  \\ 
S-AFM &  0.83        &  12.5 &  0.62   \\
AFM & 0.72          & 5.13 & 0.75 \\ 
\end{tabular}   
\end{ruledtabular}
\label{table:magnec_pr} 
\end{table}

In contrast to $\alpha$-RuCl$_3$, the OsCl$_3$ presents a FM ground state followed by the ZZ-AFM configuration, which is higher in energy by 3.0 meV/Os-atom; both magnetic phases are characterized by an energetic preference for in-plane magnetization, with a sizeable MAE of $-27.68$ (FM) and $-18.71$\,meV/Os-atom (ZZ-AFM). Our findings are in good agreement with Ref.\,\onlinecite{sheng_prb}, which reported on small energy difference between the  FM and ZZ-AFM configurations, and in-plane magnetization. Similarly to $\alpha$-RuCl$_3$, the  spin density of FM OsCl$_3$ is centered around the transition metal atoms, as can be seen in Fig.~\ref{fig:FigMagnetic}(d), and  posses a net magnetic moment of 0.9\,$\mu_{B}$/Os-atom. The FM ground state exhibits a band gap of 0.54 eV, while the ZZ-AFM configuration has a gap of 0.77 eV. 

\begin{figure}
\centering
 \includegraphics[scale=1]{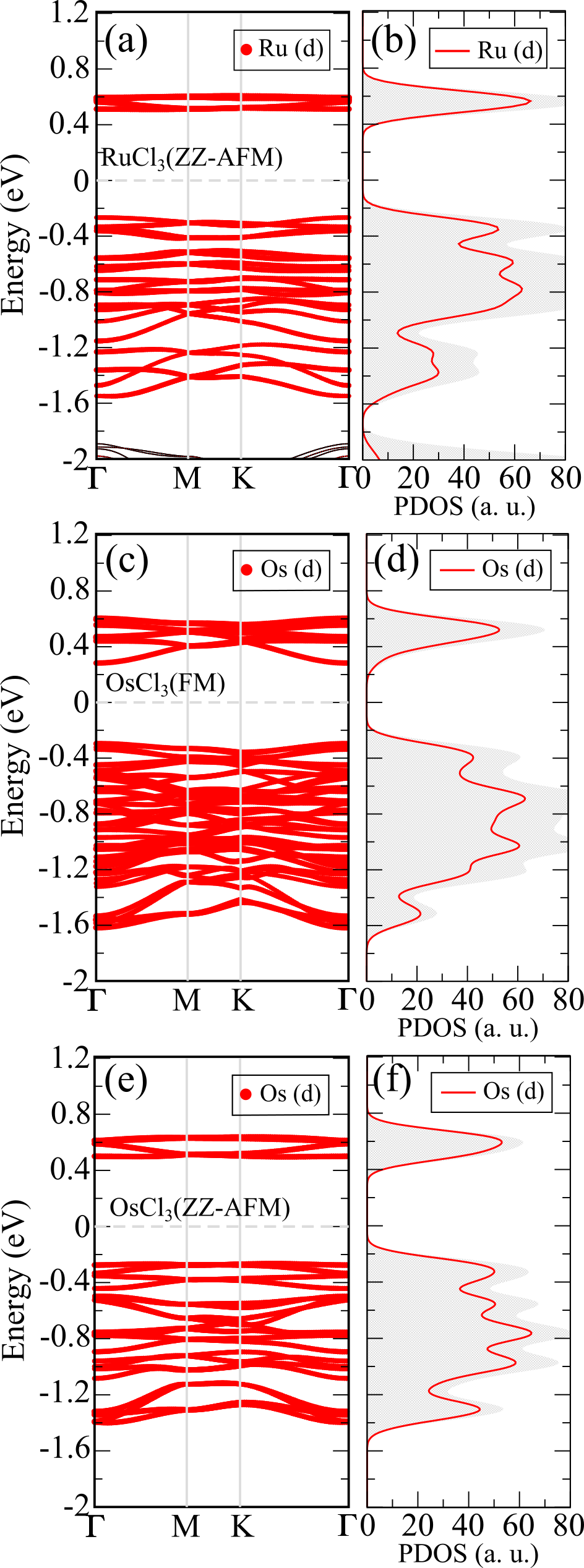}
 \caption{DFT+$U$+SOC band structures and projected density of states of the transition metal trichlorides monolayers: (a)--(b) RuCl$_3$(ZZ-AFM), (c)--(d) OsCl$_3$(FM), and (e)--(f) OsCl$_3$(ZZ-AFM). Red lines represent the contribution from Ru-4$d$ (Os-5$d$) states. The shaded regions in (b), (d) and (f) correspond to the total density of states.}
 \label{fig:bands}
\end{figure}

The Mott insulating states of MLs of $\alpha$-RuCl$_3$ and OsCl$_3$ monolayers are evidenced by the band structures and projected density of states shown in Fig.~\ref{fig:bands}. In both materials, the gap appears due to the splitting of $j_\text{eff}=1/2$ states in lower and upper Hubbard bands, which are originated from Ru(Os)-$t_{2g}$ states. Moreover, from the projected density of states, one can notice that states from $-1.6$ to $1.2$ eV are essentially formed by the Ru(Os)-$d$ states, which are gapped due to interplay of SOC the Coulomb repulsion. In the case of ML OsCl$_3$, we also show the band structure of the ZZ-AFM configuration [Fig.~\ref{fig:bands}(e)] has the same features of the RuCl$_3$(ZZ-AFM) [Fig.~\ref{fig:bands}(a)]. The ground states of $\alpha$-RuCl$_3$ and OsCl$_3$ exhibit band gaps of 0.74 and 0.54 eV (see table~\ref{table:magnec_pr}), respectively. The former is in good agreement with previous calculations using small U values,~\cite{Valenti_PRL} although photoemission studies for bulk samples indicate values of around 1.9 eV.~\cite{Sinn_Photo} It is worthy mentioning that in the case of ML OsCl$_3$, the neglecting of the Hubbard U term leads to a quantum hall anomalous insulating state with very small band gaps of $\sim$ 67 meV.~\cite{sheng_prb} 

\subsection{Effects of graphene on $\alpha$-RuCl$_3$ and OsCl$_3$} 

\begin{figure}
\centering
 \includegraphics[width=\columnwidth]{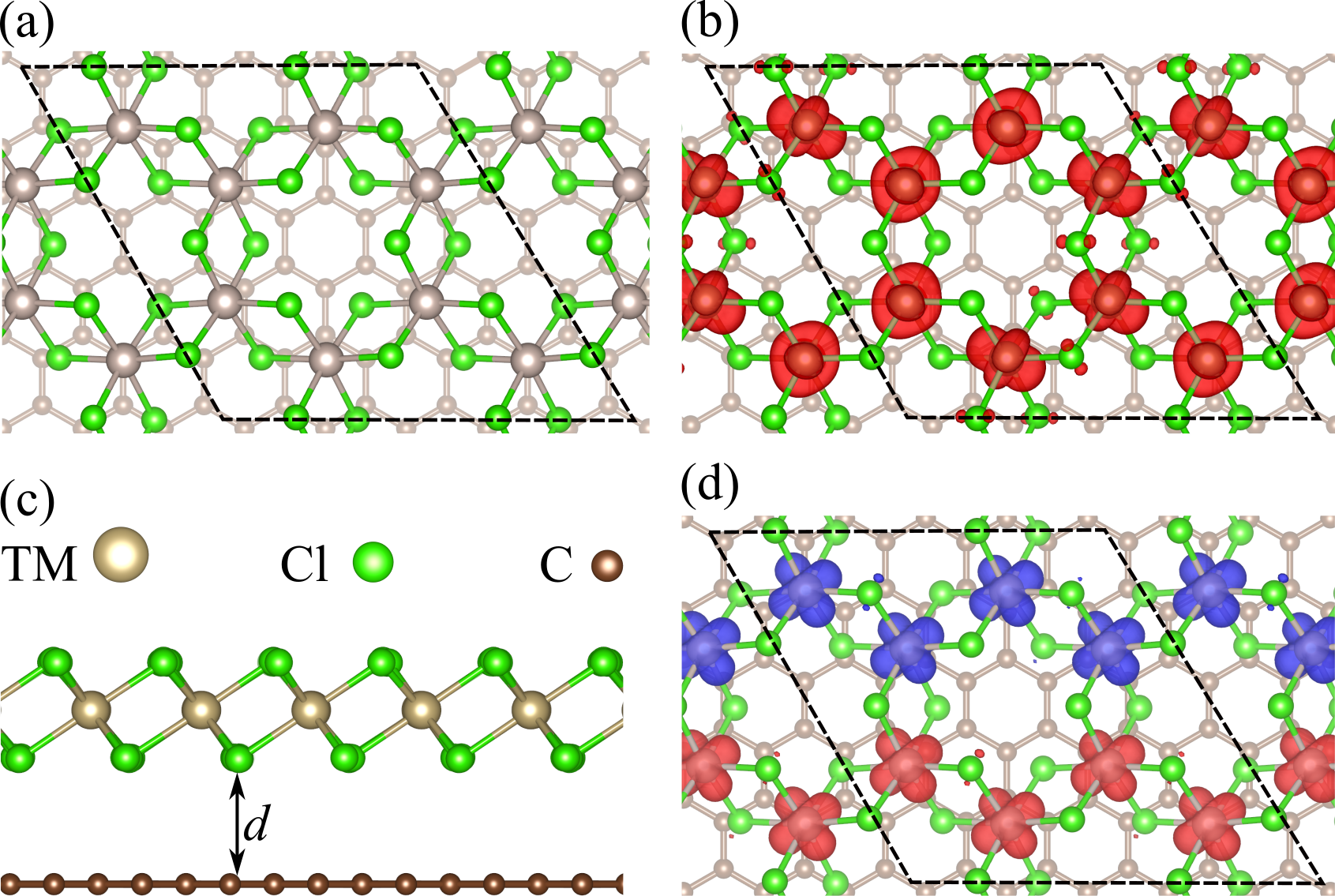}
 \caption{(a) Optimized structural model of $\alpha$-RuCl$_3$/Gr and OsCl$_3$/Gr. (b) Spin-resolved charge density isosurface corresponding to ferromagnetic phase of OsCl$_3$/Gr. Similar spin-density is shown in (d) for $\alpha$-RuCl$_3$/Gr. (c) Side view of our optmized structure, where the interlayer distance $d$ = 3.6 \AA{}.}
 \label{fig:Fig2}
\end{figure}

We now address the effects of graphene on the electronic and magnetic properties of both Ru and Os compounds. Van der Walls heterostructures have been employed as an alternative to tailor the properties of 2D materials without any drastic chemical or structural modification. For instance, graphene can be used to enhance the Kitaev interactions and the spin-split of bands in the ML of $\alpha$-RuCl$_3$.~\cite{Valenti_PRL,mashhadi_Nanolett} To investigate the energetics and electronic structure of the monolayers of $\alpha$-RuCl$_3$ and OsCl$_3$ on graphene, we perform structural optimizations of lattice parameters considering $\sqrt 3\times\sqrt 3$ hexagonal supercells, such as shown in Fig.~\ref{fig:Fig2}(a). In the optimized structures the graphene lattice parameter is expanded by 0.84 \% and 1.68 \%, when interacting with $\alpha$-RuCl$_3$ and OsCl$_3$, respectively. In contrast, in Ref.\,\onlinecite{Valenti_PRL} the authors kept the graphene lattice parameter fixed, with nearest neighbor distances of 1.42 \AA{}. 
Moreover, in our optimized structures the obtained interlayer distance between graphene and $\alpha$-RuCl$_3$ (OsCl$_3$) is found to be around 3.6 \AA{}, as illustrated in Fig.~\ref{fig:Fig2}(b). We evaluate the binding energy of both monolayers on graphene as follows,
\begin{equation}
E_b = \frac{E(TM\text{Cl}_{3}/\text{Gr}) - E(TM\text{Cl}_{3}) - E(\text{Gr})}{Area},    
\end{equation}
where $TM$\,=\,\{Ru,Os\}. $E(TM\text{Cl}_{3}/\text{Gr})$ is the total energy of our optimized structures, whereas $E(TM\text{Cl}_{3})$ denotes the total energy of monolayer of $\alpha$-RuCl$_3$ (OsCl$_3$). $E(\text{Gr})$ is the total energy of graphene sheet. We find binding energies of -17.45 meV/\AA{}$^2$ and -16.47 meV/\AA{}$^2$ for $\alpha$-RuCl$_3$/Gr and OsCl$_3$/Gr, respectively. It is important to mention that binding energy between graphene sheets are around -16.8 meV/\AA{}$^2$.\cite{BEgraphene} These results emphasize the Van der Walls nature of interactions between the transition metal trichlorides and graphene. Therefore, it is unlikely that heterostructures made of transition metal trichlorides and graphene, exhibit considerable epitaxial in-plane strain values.

\begin{figure*}
\centering
 \includegraphics[scale=1.03]{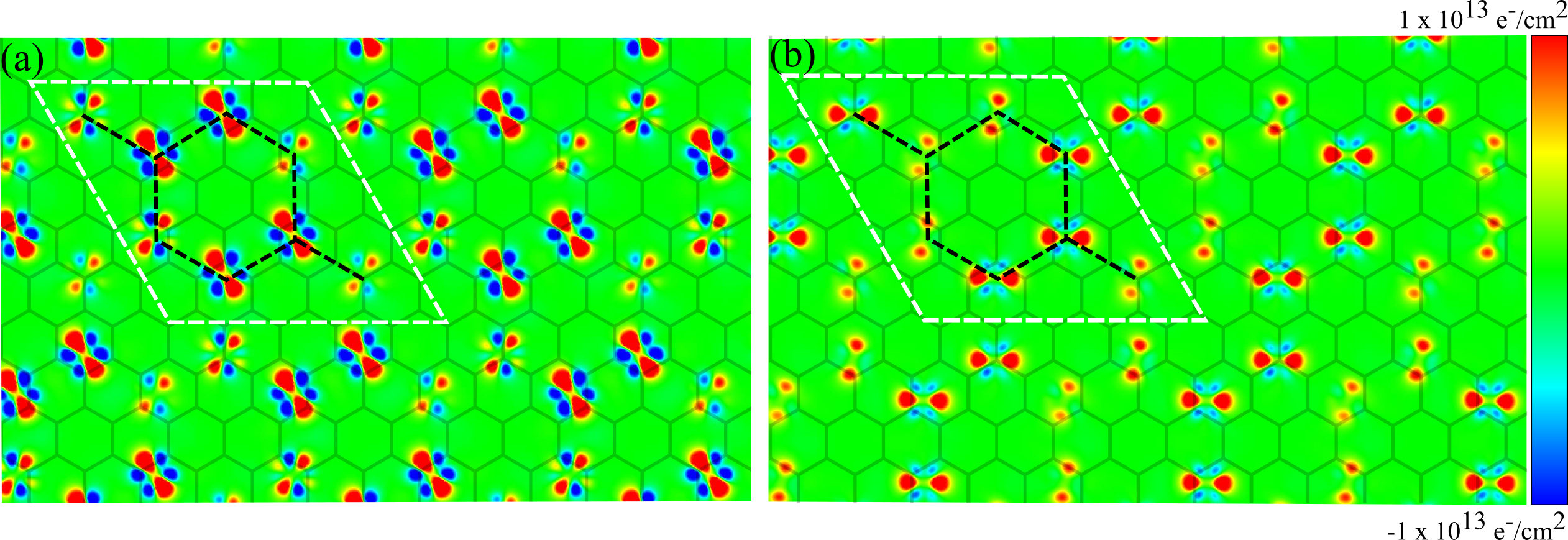}
 \caption{In plane $\Delta \rho$ cut of isosurface associated with charge transfer from graphene to (a) $\alpha$-RuCl$_3$ and (b) OsCl$_3$ monolayers. White dashed lines represent the supercells employed in our calculations. Black dashed lines show the hexagonal Ru and Os lattice to (a) $\alpha$-RuCl$_3$ and (b) OsCl$_3$.}
 \label{fig:cutcharge}
\end{figure*}
Next, we address the charge transfer between graphene and ML $\alpha$-RuCl$_3$ (OsCl$_3$) by calculating the Barder charges of the heterostructures and the isolated systems as well. Our results are displayed in Table~\ref{table:barder_charge}. According to our findings, graphene donates $3.2\times 10^{13}\,e/\text{cm}^2$ to $\alpha$-RuCl$_3$, and $2.1\times 10^{13}\,e/\text{cm}^2$  to OsCl$_3$. Roughly speaking, these results can be explained by the deep work function ($\Phi$) of the transition metal trichlorides, which are 6.10 and 5.51\,eV for  single layer $\alpha$-RuCl$_3$ and OsCl$_3$, respectively, whereas $\Phi$\,=\,4.6\,eV for graphene.\cite{rizzo_NanoLett}

\begin{table}[!htb]
\caption{Barder charges (in units of $e$/unit cell ($\sqrt{3} \times \sqrt{3}$)) of isolated $\alpha$-RuCl$_3$ (OsCl$_3$) and heterostructures with graphene. $\delta$ denotes the difference between the obtained charges of each isolated monolayer and in presence of graphene.}
\begin{ruledtabular}
\begin{tabular}{cccc}
Atom  & Charge (isolated)   & Charge (on graphene) & $\delta$ \\ \hline
             &     \multicolumn{2}{c}{$\alpha$-RuCl$_3$}   & \\ 
Ru &   53.73 &  53.88  & 0.15  \\
Cl  &  178.18  & 178.45  & 0.27  \\  
 &    \multicolumn{2}{c}{OsCl$_3$}   &        \\
 Os &  52.91     &  53.00 &  0.09   \\
Cl  &  179.01     &  179.19 &  0.18  \\ 
\end{tabular}
\end{ruledtabular}
\label{table:barder_charge} 
\end{table}

In Fig.~\ref{fig:cutcharge} we display the associated charge transfer map given by $\Delta \rho=\rho[TM\text{Cl}_3/\text{Gr}]-(\rho[TM\text{Cl}_3]+ \rho[\text{Gr}])$, with $TM$\,=\,\{Ru, Os\}. In agreement with the obtained Barder charges, we find that graphene donates electrons to the transition metal thrichlorides monolayers. Indeed, $p$-type doping of graphene in contact with $\alpha$-RuCl$_3$ has been experimentally observed, and supported by {\it first-principles} DFT calculations.\cite{mashhadi_Nanolett,gerberPRB2020ab,Valenti_PRL} More interestingly, we observe an inhomogeneous electron doping of both $\alpha$-RuCl$_3$ and OsCl$_3$, which is more pronounced in the former [Fig.~\ref{fig:cutcharge}(a)]. Such inhomogeneous doping can be explained by the different hoppings between the $TM$Cl$_3$ monolayer and the $\pi$-orbitals of the graphene sheet; which  are  sensitive to the stacking geometry at the \rucl-Gr and \oscl-Cr interfaces. These findings are consonance with the charge modulation as observed in recent Raman spectroscopy, and electronic transport measurements.\cite{zhouPRB2019evidence,Nano_modulation} It is important to mention that inhomogeneous charge doping can be deleterious to the charge carries transport in those systems, for instance, through the formation of electron-hole puddles,\cite{martinNatPhys2008,miwaAPL2011} and have important effects on the magnetic interactions between the $TM$ atoms within the distinct charge domains. It is worth noting that such an inhomogeneous net charge distribution, although less intense, has also been observed in \oscl/Gr, Fig.\,\ref{fig:cutcharge}(b).

\begin{figure*}
\centering
 \includegraphics[scale=0.95]{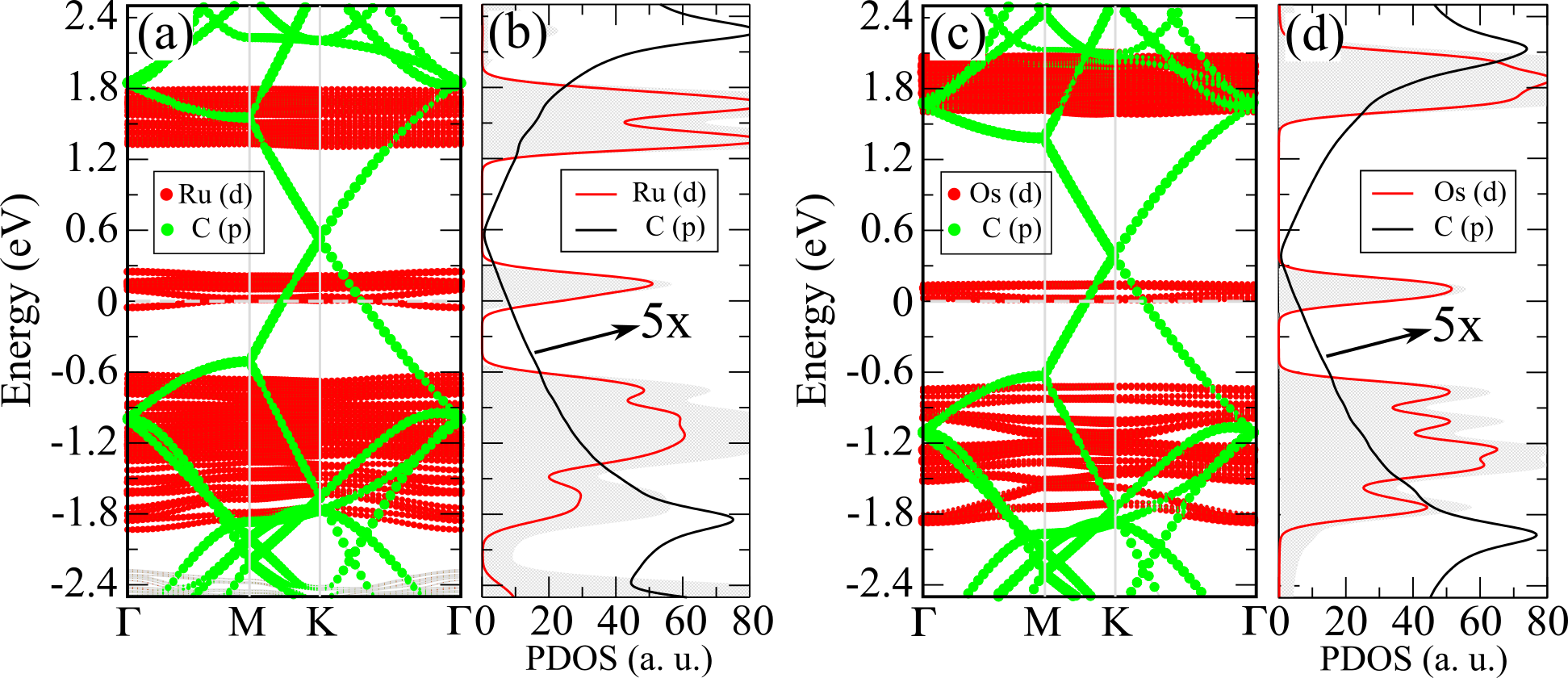}
 \caption{Orbital resolved  band structures and projected density of states of (a)--(b) $\alpha$-RuCl$_3$/Gr and (c)--(d) OsCl$_3$/Gr. The Ru-4$d$ (Os-5$d$) states are shown in red, whereas C-$2p$ states are shown in green.}
 \label{fig:bands_pdos_TMDGr}
\end{figure*}

The interaction with graphene also gives rise to important effects on the electronic and magnetic properties of the transition metal trichlorides.
As can be seen in Figs.~\ref{fig:bands_pdos_TMDGr}(a) and (c), the donated electrons occupy the narrow upper Hubbard bands (UHB) of both $\alpha$-RuCl$_3$ and OsCl$_3$, giving rise to correlated metallic phases. As a result, graphene Dirac cone appears around 0.6\,eV (0.5\,eV) above the Fermi level in $\alpha$-RuCl$_3$ (OsCl$_3$)/Gr, in good agreement with previous reports.~\cite{Valenti_PRL,rizzo_NanoLett} This also can be observed in our calculated projected density of states shown in Figs.\,\ref{fig:bands_pdos_TMDGr}(b) and (d), where one can also notice the doping of the upper Hubbard bands. Besides the (partial) occupancy of the UHB, there is also a downshift of the high energy Ru(Os)-$e_{g}$ like states in the presence of graphene. In fact, these states downshift by around 0.40 eV in $\alpha$-RuCl$_3$ and 0.36 eV in OsCl$_3$.

Focusing on the magnetic properties, we found that the ground state configuration of $\alpha$-RuCl$_3$, namely  ZZ-AFM phase  with out-of-plane magnetization, remains the same in $\alpha$-RuCl$_3$/Gr. However, the total energy difference between the  ZZ-AFM  and FM configurations increases  from  4.5 to 26.2\,meV/Ru-atom, thus indicating that the energetic preference for the ZZ-AFM  phase has been strengthened in \rucl/Gr. Meanwhile there is a reduction of the  MAE from 0.163 to 0.118\,meV/Ru-atom. On the other hand, in OsCl$_3$/Gr the ground state configuration of OsCl$_3$ changes from FM to ZZ-AFM, where the latter becomes  more stable than the former by 8.19 meV/Os-atom. In both cases the energetic preference for the in-plane magnetization is maintained, however, somewhat similar to its counterpart, \rucl/Gr, there is a reduction of the MAE, \textit{viz.}:  $-27.68\rightarrow -15.17$\,meV/Os-atom (FM phase), and $-18.71\rightarrow-14.04$\,meV/Os-atom (ZZ-AFM). Since the structural changes on the $\alpha$-RuCl$_3$ and OsCl$_3$ MLs due to their interaction with the graphene sheet are nearly negligible, we can infer that these changes (on the magnetic properties) are mostly dictated by the net charge transfer between graphene and the transition metal trichlorides. The role played by the Gr\,$\leftrightarrow$\,$TM$Cl$_3$  charge transfers will be discussed in the next subsection. 

As pointed out by Freeman and coworkers,~\cite{freeman_MAE} the magnetic anisotropy energy depends on the relative position of occupied and unoccupied $d$ energy levels, and on the coupling between them through the angular momentum operator $\mathbf{L}$. According to the authors, based on the second-order perturbation theory, the MAE can be estimated as follows, 
\begin{equation}
\text{MAE}\,\approx\,\xi^{2} \sum_{o,u} \Big[\frac{|\langle o |L_z | u \rangle|^{2}- |\langle o |L_x | u \rangle|^{2}}{\epsilon_{u}-\epsilon_{o}}\Big],
\end{equation}
for in-plane (out-of-plane) magnetization along the $x$ ($z$) direction.
 $\epsilon_{u}$ and $\epsilon_{o}$ are the eigenvalues of the corresponding eigenstates, unoccupied ($u$) and occupied ($o$). Using the equation above, combined with the DFT+U+SOC calculations, we can have a detailed orbital understanding of the MAE results upon the formation of \rucl/ and \oscl/Gr interfaces.

\begin{figure}
\centering
 \includegraphics[width=\columnwidth]{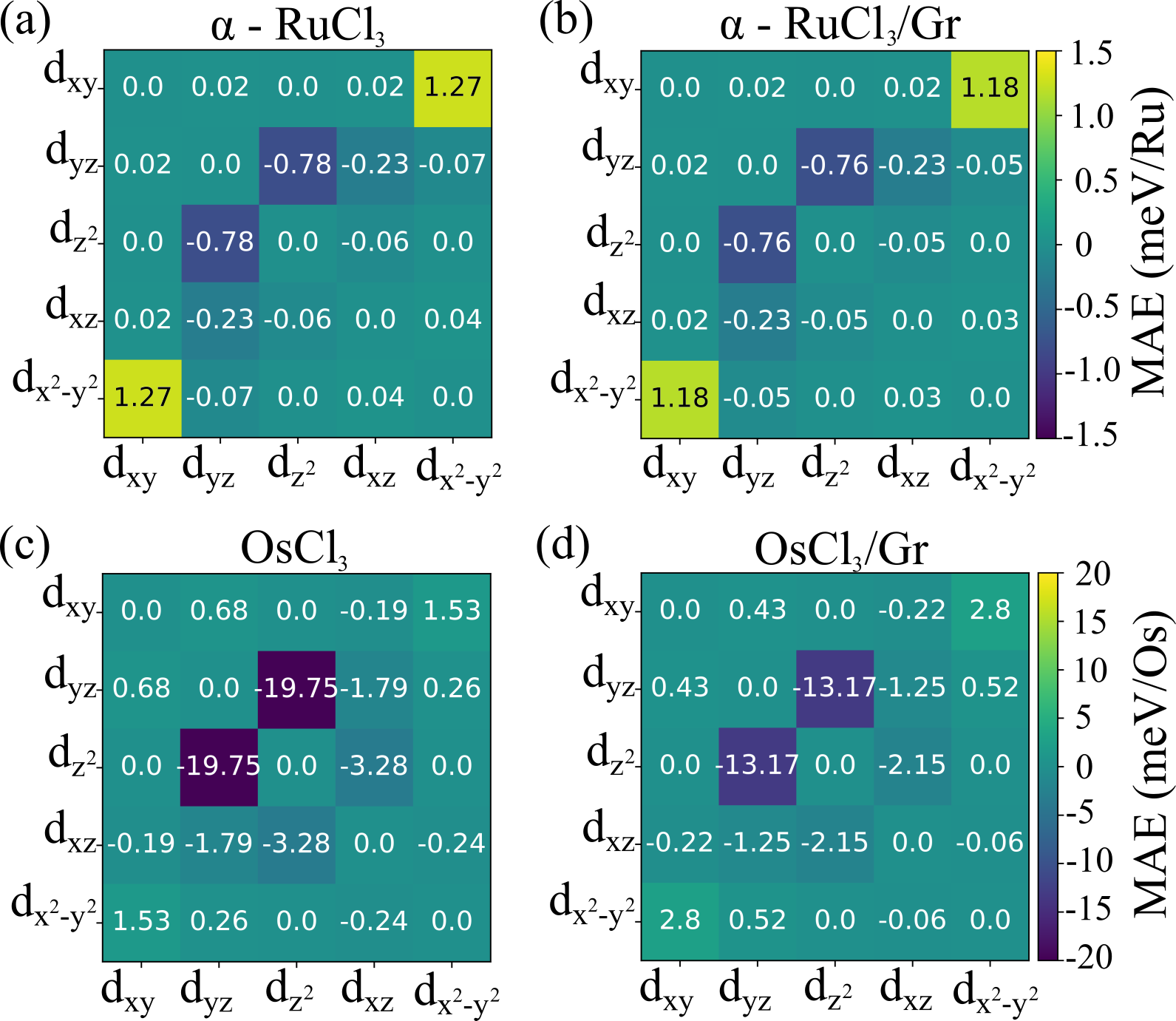}
 \caption{Orbital resolved magnetic anistropy energies (in meV/TM-atom) for (a) \rucl , (b) \rucl/Gr, (c) OsCl$_3$, and (d) \oscl/Gr. The blocks in our plots emphasize the contribution of the correspoding matrix elements.}
 \label{fig:mae}
\end{figure}

Our orbital resolved MAE results, as displayed in Fig.\,\ref{fig:mae}, reveal that the out-of-plane magnetization of $\alpha$-RuCl$_3$ is mostly dominated by the in-plane Ru-$4d$ orbitals via the matrix element $\langle\,d_{x^2-y^2}|L_z|d_{xy}\,\rangle$, but there is also a contribution from out-of-plane orbitals, $\langle\,d_{z^2}|L_x|d_{yz}\,\rangle$, favoring the in-plane magnetization. As shown in Fig.\,\ref{fig:mae}(a), the values of the former term are larger than the latter one by 0.490\,meV/Ru-atom. This energy difference between the matrix elements reduces to 0.413\,meV/Ru-atom in $\alpha$-RuCl$_3$/Gr [Fig.\,\ref{fig:mae}(b)], which is consistent with the reduction of MAE discussed above. The OsCl$_3$ and OsCl$_3$/Gr systems present a somewhat similar picture, where the matrix element $\langle\,d_{z^2}|L_x|d_{yz}\,\rangle$ dictates the energetic preference for in-plane magnetization, which in its turn reduces (by $\sim$\,6.6\,meV/Os-atom) upon the presence graphene sheet, as shown in Figs.\,\ref{fig:mae}(c) and (d).

\begin{figure*}
\centering
 \includegraphics[scale=1.0]{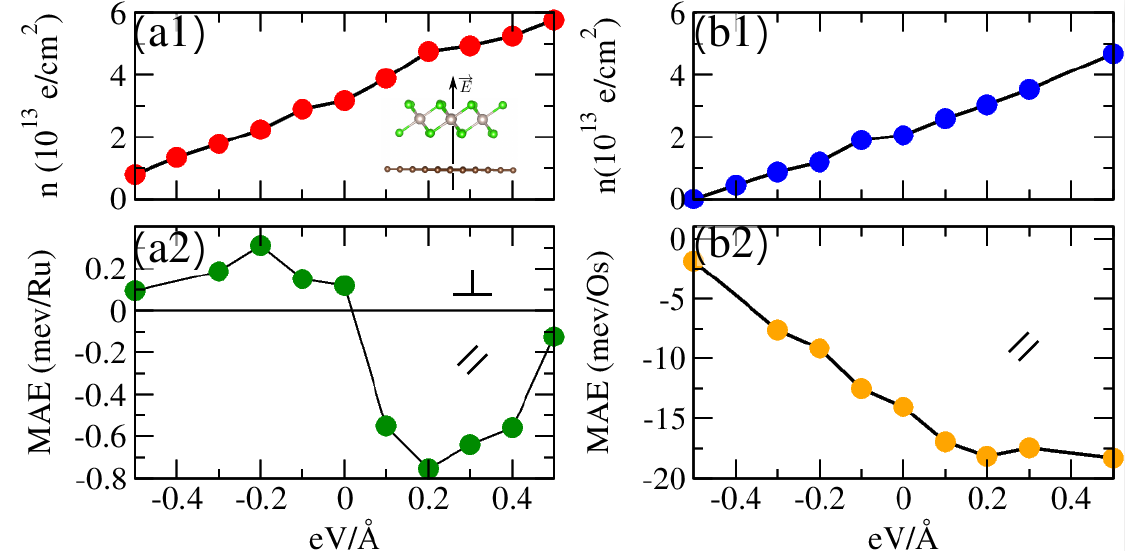}
 \caption{Electron charge transfer as a function of an external electric field, for (a1) RuCl$_3$/Gr and (b1) OsCl$_3$/Gr. The direction of the external field is illustrated in the inset of (a1). In (a2) and (b2) we show the obtained MAEs for \rucl/Gr and \oscl/Gr, respectively. The parallel ($\parallel$) and perpendicular ($\perp$) to the surface magnetizations are also indicated.}
 \label{fig:fig7_new}
\end{figure*}

\subsection{Electric field induced metal-insulator transitions and control of magnetic anisotropy}

The suitable control of the electronic and magnetic properties of materials by external agents, like mechanical pressure and electric field, is an important issue to the development of new electronic devices. For instance, metal-insulator switching in transition metal oxides and intermetallic compounds,\cite{triscone_adm, Zou_2019, rozenberg_adfm} band alignment,\cite{souza2016switchable,padilha2017,pontes2018layer} control of the magnetic phases\cite{jiang2018electric,morell2019control} in 2D materials mediated by external electric field (EEF) and mechanical pressure.\cite{song2019switching,deus2020magnetic} In the case of the $\alpha$-RuCl$_3$/ and OsCl$_3$/Gr heterostructures, in addition to the Gr\,$\rightarrow$\,$TM$Cl$_3$ electron doping (discussed above), we have also examined the effect of EEFs on the control of the net charge transfers, and on the magnetic/electronic properties as well. 

\begin{figure}
\centering
 \includegraphics[scale=0.85]{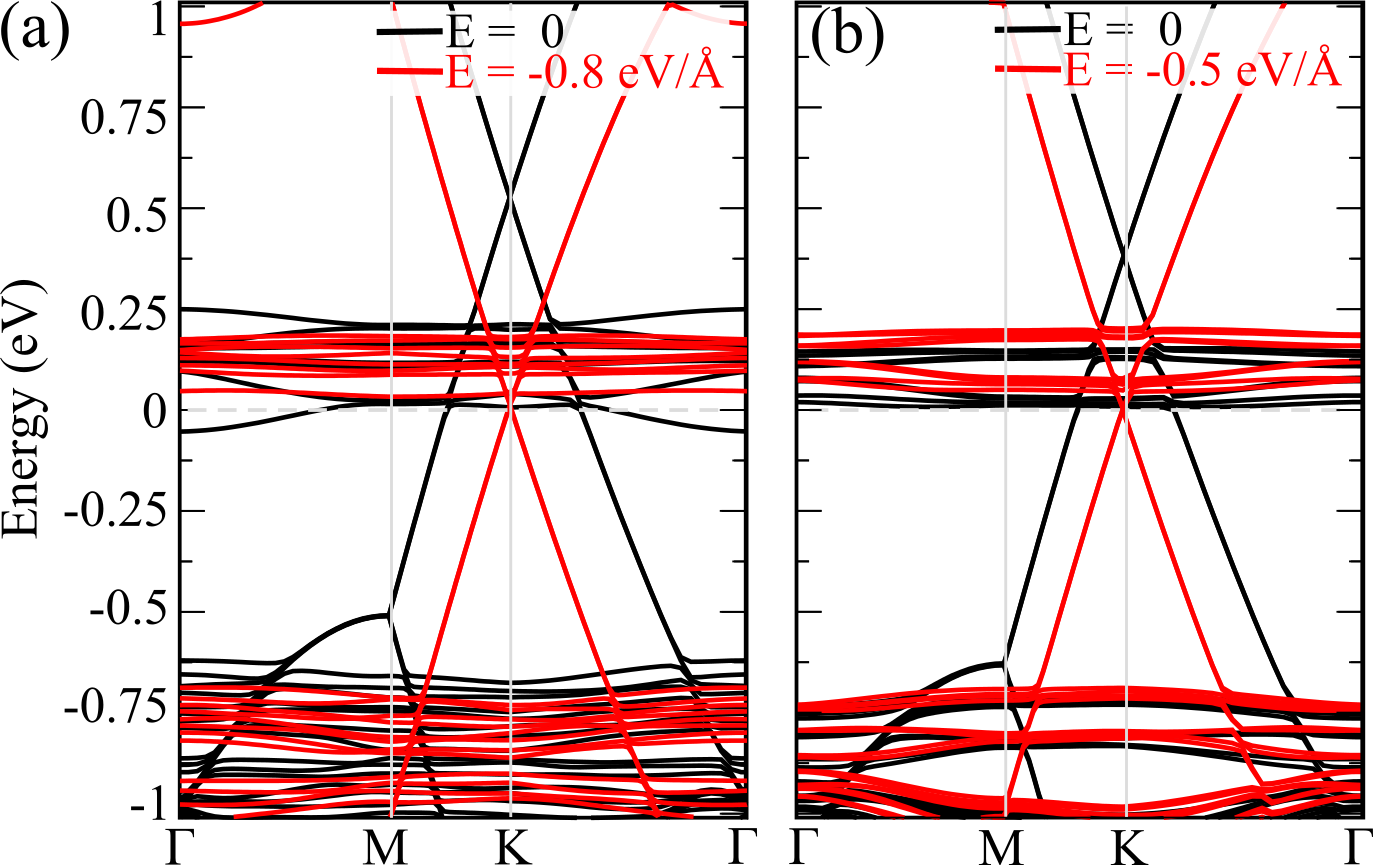}
 \caption{Band structures of (a) $\alpha$-RuCl$_3$/Gr and (b) OsCl$_3$/Gr, on the presence of an external electric fields of -0.8 eV/\AA{} and -0.5 eV/\AA{}, respectively. Bands shown in black (red) denote the calculated band structures without (with) an external electric field.}
 \label{fig:bands_E_field_OsCl}
\end{figure}

As shown in Figs.\,\ref{fig:fig7_new}(a1) and (b1),  \rucl/Gr and \oscl/Gr heterostructures present nearly linear behavior of charge transfer as a function of the EEF. The $n$-type doping of the  $TM$Cl$_3$ MLs decreases upon negative values of EEF. For an EEF of about $-0.65$\,eV/\AA\, the electron transfer from graphene to $\alpha$-RuCl$_3$ has been suppressed; likewise such a  suppression occurs for EEF of $\sim$\,$-0.50$\,eV/\AA\, in the case of OsCl$_3$/Gr. In the opposite direction, one can increase the electron doping of $TM$Cl$_3$ MLs mediated by positive values of EEF. For instance, for EEF of 0.3\,eV/\AA\, we  find an electron doping of the \rucl\, and \oscl\,MLs of  about of 5.0\,$\times$ and 3.5\,$\times 10^{13}$\,electrons/cm$^{2}$, respectively. Therefore, the occupancy of the UHB can be controlled by the EEF, as can be seen in the band structures shown in Figs.\,\ref{fig:bands_E_field_OsCl}(a) and (b). In particular, we observe insulating phases for $\alpha$-RuCl$_3$ and OsCl$_3$ upon EEFs of $-0.7$ eV/\AA\, and  $-0.5$\,eV/\AA, respectively. The suppression of the charge transfers leads also to considerable downshift of the graphene's Dirac cone, as expected. As a result, metal-insulator transitions can be induced  through external electric fields, which in its turn, controls the occupancy of the UHBs. Thus, our results demonstrate an alternative way to tune the 4$d$ and 5$d$ band filling in these compounds, and the magnetic properties of the $TM$Cl$_3$ systems. Indeed, as shown in Figs.\,\ref{fig:fig7_new}(a2) and (b2), the MAE of $TM$Cl$_3$/Gr can be tuned by the EEF.

In Fig.\,\ref{fig:fig7_new}(a2) we present the MAE of RuCl$_3$/Gr as a function of the EEF. It is noticeable that, (i) within $|\text{EEF}|$\,$\leq$\,0.5\,eV/\AA,  the strength of the out-of-plane magnetization increases from 0.118\,meV/Ru-atom (EEF\,=\,0) to 0.311\,meV/Ru-atom for an EEF of $-0.2$\,eV/\AA, corresponding to a reduction of the Gr\,$\rightarrow$\,RuCl$_3$ charge transfer, $\Delta\rho$, from 3.2\,$\times$ to 2.2\,$\times$\,10$^{13}\,e/\text{cm}^2$. In contrast, (ii)  the in-plane magnetization becomes energetically more favorable for positive values of the EEF, where we found MAE of $-0.754$\,meV/Ru-atom for an EEF of 0.2\,eV/\AA. In this case, the $n$-type doping of RuCl$_3$ increases, with $\Delta\rho$ of  4.8\,$\times$\,10$^{13}\,e/\text{cm}^2$. 

We can gain further understanding of the role played by the EEF by analysing its effect on the (DFT+$U$) orbital contribution to the MAE, eq.\,(2). 
Overall, we find that the orbital contributions to the MAE are reduced in comparison with those with no EEF. For instance, the out-of-plane contribution given by the matrix element $\langle\,d_{x^2-y^2}\,|L_z|\,d_{xy}\,\rangle$ reduces from 1.176 to 0.891\,meV/Ru-atom, while the in-plane contribution via $\langle\,d_{z^2}\,|L_x|\,d_{yz}\,\rangle$ reduces (in absolute values) from 0.763 to 0.445\,meV/Ru-atom. The larger energy reduction of the latter matrix element leads to an energetic preference for the out-of-plane magnetization, as discussed in (i). On the other hand, for EEF of 0.2\,eV/\AA\, [(ii)], the preference for in-plane  magnetization, MAE\,=\,$-0.754$\,meV/Ru-atom, is mostly ruled by the matriz element $\langle\,d_{x^2-y^2}\,|L_z|\,d_{xy}\,\rangle$ integrated over orbitals with opposite spins.~\cite{freeman_MAE,yangPRB2017strain}

Meanwhile, the  energetic preference for in-plane magnetization in OsCl$_3$/Gr is maintained within $|\text{EEF}|\leq 0.5$\,eV/\AA, Fig.\,\ref{fig:fig7_new}(b2). There is an increase of the MAE from $-14.04$\,meV/Os-atom (EEF\,=\,0) to  $-18.16$\,meV/Os-atom  for EEF\,=\,0.2\,eV/\AA; in the opposite direction we  find a nearly linear reduction of the MAE,$-14.04$\,$\rightarrow$\,$-1.84$\,meV/Os-atom, for negative values of EEF from 0 to $-0.5$\,eV/\AA. In the latter limit, the net charge transfer from graphene to  OsCl$_3$ is suppressed, $\Delta\rho=0$, suggesting that the preferential  magnetization can be tuned from in-plane to out-of-plane mediated by higher values of EEF or $p$-type doping of the OsCl$_3$ ML. We have also examined the effect of EEF on the MAE in light of the perturbation theory. The calculated  matrix elements, present in eq.\,(2), for EEF\,=\,0.2 and $-0.5$\,eV/\AA, respectively, where we show that the MAE's changes as a function of the EEF  are mostly ruled by the matrix elements $\langle\,d_{z^2}\,|L_x|\,d_{yz}\,\rangle$, and $\langle\,d_{z^2}\,|L_x|\,d_{xz}\,\rangle$. For instance,  contribution from the former element increases/decreases from $-13.165$ (EFF=0) to $-15.139$/$-3.99$\,meV/Os-atom for a EEF of 0.2/$-0.5$\,eV/\AA. 

It is worth noting that the changes on the MAE as a function of the EEF are dictated by the occupation of the Ru-$4d$ and Os-$5d$ orbitals,  UHBs resonant or nearly resonant to the Dirac point. Thus, in order to stress the  importance of the UHBs  on the magnetization direction, we calculated the MAE of $n$-type ($p$-type) doped  \rucl\, (\oscl) ML. We found a magnetic transition from out-of-plane to in-plane, MAE\,=\,0.163\,$\rightarrow$\,$-0.213$\,meV/Ru-atom for $n$\,=\,1\,$e$, in accordance with the energetic preference for the in-plane magnetization of  \rucl/Gr upon EEF\,$>$\,0 [Fig.\,\ref{fig:fig7_new}(a2)]. 
Meanwhile,  we find MAE\,=\,$-16.41$\,meV/Os-atom in  the  $p$-type doped ($p$\,=\,0.033\,$h$ upon an EEF of $-0.7$\,eV/\AA) OsCl$_3$ ML,  confirming the tendency of change on the magnetic orientation, in-plane\,$\rightarrow$\,out-of-plane, for EEF\,$<$\,$-0.5$\,eV/\AA\, [Fig.\,\ref{fig:fig7_new}(b2)]. Thus, our findings based on electron and hole doping of \rucl\, and \oscl\, MLs, respectively, provide strong evidence that the occupation of the UHBs, indeed, play an important role on the tuning of the MAE.

\section{Summary and Conclusions}

In summary, we performed DFT+$U$+SOC calculations to investigate the effects of graphene on the electronic and magnetic properties of monolayers of $\alpha$-RuCl$_3$ and OsCl$_3$, \textit{i.e.} \rucl/Gr and \oscl/Gr heterostructures. We find that the $\alpha$-RuCl$_3$ and OsCl$_3$ MLs  become $n$-type doped, with $n$ of the  order of $10^{13}\,e/\text{cm}^2$, characterized by an inhomogeneous spatial distribution of the net charge density, which depend on the stacking geometry (orbital hopping) between Ru(Os) and carbon atoms. The corresponding charge transfer gives rise to correlated metallic phases in both materials, mediated by the  partially occupied upper Hubbard bands.  We demonstrate that metal-insulator transitions can be induced in \rucl/ and \oscl/Gr using external electric fields, which in turn controls the occupancy of the Ru-4$d$ Os-5$d$. More important, such control on the occupancies of the upper Hubbard bands leads to tuneable magnetic properties. We found that the in-plane magnetization becomes energetically favorable than the out-of-plane one in $\alpha$-RuCl$_3$ upon $n$-type doping. Meanwhile, in the opposite direction, $p$-type doping of OsCl$_3$ results in an in-plane\,$\rightarrow$\,out-of-plane transition in the preferential magnetization direction. Our findings suggest that the occupancies of the Hubbard bands of the transition metal thriclorides are the key ingredients to tune the magnetic easy-axis as well as the magnetic anisotropy energies in these two-dimensional correlated materials. This tunning can be achieved with graphene and external electric fields.

\begin{acknowledgments}

The authors acknowledge financial support from the Brazilian agencies 
CAPES, CNPq, FAPEMIG,  and the National Laboratory for Scientific Computing (LNCC/MCTI, Brazil, project SCAFMat2) for providing HPC resources of the SDumont supercomputer, which have contributed to the research results, URL: http://sdumont.lncc.br.

\end{acknowledgments}

\bibliography{ruoscl3_v1.bib}

\end{document}